\documentclass[12pt]{article}
\usepackage{natbib}
\usepackage[usenames,dvips]{color}
\usepackage{graphicx}
\begin{document}
\begin{flushright}
{\underline{\bf To appear in Journal of Astrophysics and Astronomy}}
\end{flushright}

\begin{center}
{\bf \large
An upper limit on the ratio between the Extreme
Ultraviolet and the bolometric luminosities of stars hosting
habitable planets} \\
\vskip0.5in
{\large Sujan Sengupta} \\ \vskip0.1in 
Indian Institute of Astrophysics, Koramangala 2nd Block, Bangalore 560034,
India; sujan@iiap.res.in
\end{center}

\vskip0.5in

\begin{abstract}
A large number of terrestrial planets in the classical habitable
zone of stars of different spectral types  has already been
discovered and many are expected to be discovered in near future.
However, owing to the lack of knowledge on the atmospheric
properties, the ambient environment of such planets are unknown.
It is known that sufficient amount of Extreme Ultraviolet
(EUV) radiation from the star can drive hydrodynamic outflow
of hydrogen that may drag heavier species from the atmosphere of
the planet. If the rate of mass loss is sufficiently high then
substantial amount of volatiles would
escape causing the planet to become uninhabitable. Considering
energy-limited hydrodynamical mass loss with an escape rate that
causes oxygen to escape alongwith hydrogen,
I present an upper limit for the ratio between the EUV
and the bolometric luminosities of stars which constrains the
habitability of planets around them. Application of the
limit to planet-hosting stars with known EUV luminosities implies
that many M-type of stars should not have habitable planets
around them.  

\end{abstract}

\clearpage

\section{INTRODUCTION}

Two decades after the first confirmed discovery of
planets outside the solar system \citep{wolszczan94,mayor95},
we know  more than 1500 confirmed planets
of different mass, size and surface temperature that are
orbiting around stars of different spectral types.  
Many of the gas giant planets discovered are
orbiting so close to their parent stars that tidal effects of
the star and atmospheric erosion due to strong stellar
irradiation play dominant role in determining their physical
properties and evolution \citep{lammer03,baraffe04,
hubbard07,erkaev07,penz08,sanz11}. On the other
hand, many small and possibly rocky planets are recently
discovered that may have surface temperature similar to that
of the Earth. Therefore, the focus has rapidly changed into
detecting planets that may have favorable environment to
harbor life.

Classically, a habitable zone  is defined as the one that has
favorable ambient temperature to keep water in liquid state 
\citep{huang59,
hart78, kasting93, selsis07}. In the recent years
a good number of planets are detected within the habitable zone
around stars of various spectral types \citep{udry07,vogt10,pepe11,borucki11,
borucki12,bonfils13}.

It is known that about 75\% of the stars in the
extended solar neighborhood are M dwarfs \citep{reid02}.
Analysis of data obtained by using the transit method indicates
the occurence rate of small and potentially rocky planets with radius ranging
from 0.5 to 2.0 $R_{\oplus}$ ($R_{\oplus}$ is the radius of the
Earth) around M dwarf stars is about 0.51 per star
\citep{dressing13, koppa13} while that by using
the radial velocity method is 0.41 per star \citep{bonfils13}.
It is believed that a good fraction of these rocky planets are situated
within the habitable zone of their star. Therefore, there may exist a large
number of habitable planets in the solar neighborhood and
in our galaxy.

However, a habitable planet should have appropriate 
amount of hydrogen, nitrogen, oxygen gas
and water molecules that support a habitable environment.
Various thermal and non-thermal mechanisms cause hydrogen to
escape the atmosphere of a terrestrial planet. Thermal mechanisms
include hydrodynamic escape and Jeans escape. The loss of
hydrogen from a planetary atmosphere is limited either at the
homopause by  diffusion or at the exobase by energy. Diffusion
causes substantial hydrogen loss during the early evolutionary
period of a terrestrial planet.
 In today's Earth, the atmosphere is collisionless above the
tropopause and hence the barometric laws break down. 
Therefore, Jeans law is not applicable.
 Jean's law is also not applicable if the atmosphere is not
under hydrostatic equilibrium. This may occur by the 
absorption of stellar Extreme Ultraviolet (EUV) radiation 
with wavelengths ranging from 100 to 920 $\AA $. Strong EUV
irradiation heats up the hydrogen rich 
thermosphere so significantly that the internal energy of the 
gas becomes greater than the gravitational potential energy.
 This leads to the expansion of the atmosphere and powers
hydrodynamic escape of hydrogen which can drag off heavier
elements if the escape flux is high enough.  
 The Earth does not receive
sufficiently strong EUV irradiation from the Sun at present. Also
the total hydrogen mixing ratio in the present stratosphere is
as small as $10^{-5}$. Therefore, in today's Earth,
hydrogen escape is limited by diffusion through the homopause.

    The bolometric luminosity is one of the important observable stellar 
properties that  determines the distance of the Habitable Zone (HZ) or the
circumstellar region in which a planet can have appropriate
temperature to sustain water in liquid state.
On the other hand, the EUV luminosity determines the rate at
which hydrogen escapes from the atmosphere of a planet in
the HZ. If the bolometric luminosity of the star is low, the
HZ is closer to the star and hence a planet in the HZ is
exposed to stronger EUV irradiation.
 If the EUV luminosity of such faint
stars is sufficiently high then a habitable planet would lose
substantial amount of hydrogen and heavier gases. It
would lead to oxidation of the surface and accumulation of oxygen 
in the atmosphere, scarcity of hydrocarbon and significant 
reduction in surface water. All these would
make the planet uninhabitable. If a habitable planet undergoes
runway greenhouse, water vapor would reach the stratosphere 
where it would get photolyzed and subsequently should escape
the atmosphere.

  In this paper, I present an analytical expression for the
upper limit of the ratio between the EUV and the bolometric
luminosities of a planet-hosting star of any spectral type which
serves as an essential condition to ensure the presence of
sufficient amount of hydrogen, nitrogen, oxygen, water etc. that
supports the habitability of a rocky planet in the HZ of the
star. Since both $L_{EUV}$ and $L_b$ are observable quantities, 
it will help to determine if a planet in the habitable zone of a 
star is really habitable or not. Thus the limit should serve as a ready
reckoner for eliminating candidate habitable planets.

\section{CRITERIA FOR PLANETARY HABITABILITY}

The distance of a habitable planet from its
parent star is given by \citep{kasting93,koppa13a}
\begin{eqnarray}
d=\left(\frac{L_B}{L_\odot S_{eff}}\right)^{1/2}d_E
\label{hzd}
\end{eqnarray}
where $L_B$ and $L_\odot$ are the bolometric luminosity 
of the star and the Sun respectively, $d_E$ is the distance between the Earth
and the Sun (1 AU) and $S_{eff}$ is the normalized effective stellar flux
incident on the planet. 

Let $L_{EUV}$ be the EUV luminosity at the surface of a star. If
$\epsilon$ is the efficiency at which the EUV is absorbed by
the planetary atmosphere then the sphere-averaged and
efficiency corrected heating rate in the planetary
thermosphere due to stellar EUV irradiation is
\begin{eqnarray}
S=\frac{L_{EUV}}{L_B}\left(\frac{\epsilon S_{eff} L_\odot}{4\pi d^2_E}\right).
\label{ratio}
\end{eqnarray}
The heating efficiency is less than one because
part of the total incident EUV energy drives ionization,
dissociation and other reactions without striping out the atoms
or molecules from the atmosphere. 

In order to remain habitable, $S$ must be low enough so that
the rate of hydrogen escape from the planetary surface due to
EUV irradiation does not exceed a critical value that
may cause the heavier constituents including oxygen to be
entrained with the outflow of hydrogen. If the mixing ratio
of hydrogen is low in the lower atmosphere, hydrogen escape
should be limited by diffusion. Therefore the critical rate
of energy limited loss should  be comparable to that of diffusion
limited escape under such circumstance.  

 Let $S_c$ be the value of
$S$ corresponding to the critical rate of hydrogen-loss.
Therefore, the necessary condition to prevent 
hydrogen loss at critical rate by EUV irradiation can
be written as :
\begin{eqnarray}
\frac{L_{EUV}}{L_{B}}< \frac{4\pi d_E^2}{\epsilon S_{eff} L_\odot}S_c.
\label{cond}
\end{eqnarray}
Now we need to derive $S_c$, the sphere-averaged,
efficiency-corrected and energy-limited critical EUV flux that
causes hydrogen to escape at the critical rate. 

\section{ENERGY-LIMITED EUV FLUX}

 In order to derive the energy-limited critical EUV flux $S_c$,
the formalism given by  \cite{watson81} is adopted.
\cite{watson81} applied this formalism  to estimate 
the mass loss of Earth's hydrogen exosphere due to solar EUV 
irradiation. It was also used by \cite{lammer03} for explaining
the observed extended atmosphere for the
close-in hot and giant transiting planet HD 209458b
\citep{vidal03}. The formalism is derived from the
usual steady state equations of mass momentum and energy
conservation of a dynamically expanding, non-viscous gas of
constant molecular weight in which the pressure is isotropic.
Unlike the Jean's treatment, this formalism can be applied to
a dense thermosphere of hydrogen with a fixed temperature at
the lower boundary located sufficiently above the homopause
such that the mixing ratios of heavier gases are negligible as
compared to that of hydrogen. All the EUV energy is absorbed in
a narrow region with visible optical depth less than unity
and no EUV energy is available bellow this region due to
complete absorption.
The rate of mass loss estimated by using this method is within
a factor of few \citep{hubbard07} of the same provided by other
models including those that involve detailed numerical solutions
\citep{baraffe04,tian05,yelle04}. However, \cite{murray09} argued
that for the irradiated EUV flux greater than $10^4$ erg 
cm $^{-2}$ s$^{-1}$, mass loss ceases to be energy limited and
becomes radiation/recombination limited. As a consequence the 
formalism prescribed by \cite{watson81} is not applicable in that
case. 

The two equations that provide the maximum rate of
hydrogen escape and the expansion of the atmosphere
due to EUV heating are given as \citep{watson81} :
\begin{eqnarray}
\xi=\frac{2}{q+1}\left[\frac{(\lambda_1/2)^{(1+q)/2}+1}{
\lambda_0-\lambda_1}\right]^2
\label{watson1}
\end{eqnarray}
and
\begin{eqnarray}
\beta=\xi\lambda_1^2\left[\lambda_0-\left\{\frac{2}{(1+q)\xi}\right\}^{1/2}\right].
\label{watson2}
\end{eqnarray}
In the above equations the dimensionless parameters $\xi$, $\beta$,
$\lambda_{0}$ and $\lambda_{1}$ are related with the physical parameters by  
\begin{eqnarray}
\xi=F_m\left(\frac{k^2T_0}{\kappa_0GM_Pm_H}\right),
\label{xi}
\end{eqnarray}
\begin{eqnarray}
\beta=S\left(\frac{GM_Pm_H}{\kappa_0 k T^2_0}\right),
\label{beta}
\end{eqnarray}
\begin{eqnarray}
 \lambda_{0,1}=\frac{GM_Pm_H}{k T_0 r_{0,1}},
\label{radius}
\end{eqnarray}
where $r_0=R_P$ is the radius of the planet where the visible
optical depth is one and the temperature is $T_0$,
 $r_1$ is the radius of the region
 where all the EUV energy is absorbed. The optical depth of the atmosphere
to EUV energy at the lower boundary $r_0$ is much greater
than unity. $F_m$ is the sphere averaged
flux of escaping particles (escape flux of particles per
steradian per second), $k$ is the Boltzmann constant, $M_P$ is
the total mass of the planet, $m_H$ is the mass of hydrogen 
atom and $\kappa$ is the thermal conductivity of the gas which is
parameterized by $\kappa=\kappa_0\tau^q$ where $\tau$ is
the visible optical depth of the thermosphere such that
$\tau(\lambda_0)=1$.

\section{ANALYTICAL AND NUMERICAL SOLUTIONS}

 Equation~(\ref{watson1}) and equation~(\ref{watson2}) need to be
solved numerically for an arbitrary value of $q$. However,
for $q=1$ we obtain analytical solutions for $\lambda_1$ and
$\beta$ and hence $S$. Thus for $q=1$, equation~(\ref{watson1})
gives
\begin{eqnarray}
\lambda_1=\frac{2(\xi^{1/2}\lambda_0-1)}{1+2\xi^{1/2}}.
\label{sol1}
\end{eqnarray}
Substituting equation~(\ref{sol1}) in equation~(\ref{watson2}), 
 equation~(\ref{beta}) gives,
\begin{eqnarray}
S(q=1)=4\xi\left(\frac{\xi^{1/2}\lambda_0-1}{1+2\xi^{1/2}}
\right)^2\left(\lambda_0-\frac{1}{\xi^{1/2}}\right)
\left(\frac{\kappa_0 k T^2_0}{GM_Pm_H}\right).
\label{sol2}
\end{eqnarray}
Now, $S=S_c$ when $F_m=F_c$ in the expression for
$\xi$ given in equation~(\ref{xi}) where $F_c$ is the critical
rate of hydrogen escape. Hydrogen diffuses before it reaches the
thermobase if the background gas of heavier species is static.
This happens if the heavier species are  not absorbed or
cannot escape at the surface. The diffusion limit for hydrogen
is achieved when the heavier gases attain the
maximum upward velocity such that the background becomes 
non-static \citep{hunten73}. Therefore if $F_c$ is greater than or
equal to the diffusion limit, heavier species would escape. 
The rate of hydrogen loss limited by diffusion is given by 
\cite{hunten73, zahnle90}
\begin{eqnarray}
F_H(diffusion)=\frac{bg(m_s-m_H)}{kT_0(1+X_s/X_H)}
\end{eqnarray}
where $b=4.8\times10^{17}(T/K)^{0.75}$ cm$^{-1}$s$^{-1}$ is
the binary diffusion coefficient for the two species 
\citep{zahnle86}, $X_H$ and $X_s$ are the molar mixing ratio at
the exobase for hydrogen and the heavier atom with mass $m_s$
respectively and $g$ is the surface gravity of the planet.
Therefore we set
\begin{eqnarray}
F_m=F_c=R_P^2\frac{bg(m_s-m_H)}{kT_0(1+X_s/X_H)}{\rm
 particles~sr^{-1}s^{-1}}
\label{fm}
\end{eqnarray}
which gives the cross over mass $m_s=m_H+(kT_0F_c/R_P^2)/(bgX_H)$
 such that any species with mass less
than or equal to $m_s$ would be efficiently dragged by the
escaping hydrogen \citep{hunten87}. In the present derivation
I consider $m_s=m_O=16m_H$ such that the critical rate of
hydrogen loss would enable oxygen and other atoms lighter
than oxygen to be dragged with the escaping gas. I assume that
both hydrogen and oxygen are atomic near the exobase due to fast
dissociation of the molecules by photolysis \citep{murray09}.   

Therefore, from  equation~(\ref{cond}) we obtain, for $q=1$
\begin{eqnarray}
\frac{L_{EUV}}{L_B}< \frac{16\pi d_E^2\kappa_0 k T^2_0}{\epsilon GM_Pm_H
(1-e^2)^{1/2}S_{eff}L_\odot}
\frac{\xi^{1/2}(\lambda_0\xi^{1/2}-1)^3}{(2\xi^{1/2}+1)^2},
\label{result}
\end{eqnarray}
where
$\lambda_0$, the parameter for the planetary radius is given by
equation~(\ref{radius}) and
\begin{eqnarray}
\xi=R_P^2\frac{bg(m_O-m_H)}{1+X_O/X_H}\frac{k}{\kappa_0GM_Pm_H}.
\end{eqnarray}
The term $(1-e^2)^{1/2}$ in the denominator is
introduced in order to include elliptical orbit with 
eccentricity $e$.

The above analytical expression is derived by taking $q=1$. However, for a
 neutral gas, $q\simeq 0.7$ \citep{banks73}.
Therefore, I solve equation~(\ref{watson1}) and
equation~(\ref{watson2})
numerically for different values of $q\leq 1.0$ keeping all other
parameters fixed. Figure~1 presents the values of $S_c$ for
different values of $q$ and shows that $S_c (q=1) < S_c (q<1)$.
Therefore, equation~(\ref{cond}) implies that the inequality
given by equation~(\ref{result}) is valid for any value of
$q\leq 1$.
It is worth mentioning that the bolometric luminosity of
a main sequence star increases with time. For a solar-type star,
the bolometric luminosity increases by about 10 \% in every
1 Gyr. The Sun was about 30\% fainter in the visible
light during the first billion years after its birth. On the other hand the EUV luminosity is governed
by the coronal activities of a star. Since a star rotates faster
during younger age resulting into greater coronal activities,
the EUV emission rate or the EUV luminosity was higher in the
past and it decreases with time. Consequently, the rate of mass
loss from the planetary surface was higher in the past.
Since, the value of $L_{EUV}/L_{B}$ in the past was greater than
its present value, the condition derived here is valid for a
constant rate of mass loss corresponding to the present value of
$L_{EUV}/L_B$. A time-dependent solution would have provided more
stringent condition.

\section{APPLICATION TO HABITABLE EXOPLANETS }

The thermal conductivity of hydrogen $\kappa_0=4.45\times10^4$
ergs cm$^{-1}$s$^{-1}$K$^{-1}$ \citep{hanley70}. The size and the
mass of the planets are usually determined from the observed 
parameters of the planet. Transit method provides the size and the
orbital inclination angle of the planet and radial velocity method provides
the projected mass of the planet for known stellar mass.
Therefore, the two unknown parameters in equation~(\ref{result})
are $T_0$ and $S_{eff}$. 

 The lower boundary $r_0$ is located some distance
above the homopause and the temperature $T_0$ at the lower boundary may be
fixed at the equlibrium temperature of the planet \citep{watson81}. However, 
as shown in Figure~2, the above limit is weakly dependent on $T_0$ because 
the parameter $\lambda_0$ is inversely proportional to $T_0$. Therefore, 
without loss of generality, $T_0$ can be fixed at 254 K, the equilibrium 
temperature of the Earth. This makes the above expression independent of
planetary albedo because derivation of equilibrium temperature requires the
value of planetary albedo. With $T_0=254$K and taking 
$X_O=1/3$ and $X_H=2/3$, I
obtain $F_c=1.43\times10^{13}$ cm$^{-2}$ s$^{-1}$ which is
two orders of magnitude higher than that calculated by 
\cite{watson81} for the Earth irradiated by solar EUV radiation.  
This ensures the cross over mass $m_s$ is equal to $m_O$.  
The total mass of hydrogen ($M_H$) in the
atmosphere, Ocean and in the crust of the Earth is about 
$1.9\times10^{23}$ gm \citep{anders77, sharp09}. 
Therefore at this critical rate, an Earth like planet would 
lose all of its surface hydrogen in about 50 Myr. Water 
molecules would be photo-dissociated into hydrogen and oxygen
atoms that would subsequently escape the planet. Even if the
amount of water is not reduced sufficiently, such a significant
loss of hydrogen and other heavier gases such as nitrogen and
oxygen would make the environment drastically different than 
that of the present Earth.

  \cite{koppa13a} have presented relationships between the stellar effective
temperature $T_{eff}$ and the normalized stellar flux $S_{eff}$ impinging on 
the top of the atmosphere of an Earth-like planet at the habitable zone.     
These relationships are applicable in the range 2600 K$\leq T_{eff} \leq $
7200 K. According to these relationships, $S_{eff}$ is minimum at 
$T_{eff}=2600$ K and
increases with the increase in $T_{eff}$. $S_{eff}$ also varies for different
atmospheric conditions that determines the inner and the outer 
boundaries of the habitable zone. While condition like  "Recent
Venus", "Runway Greenhouse" or "Moist Greenhouse" determines the inner 
boundary, condition such as "Maximum Greenhouse" or
"Early Mars" determines the outer boundary of the habitable zone. Since,
$L_{EUV}/L_B$ in equation~(\ref{result}) is inversely proportional to
$S_{eff}$, I use the minimum value of $S_{eff}$
corresponding to $T_{eff}=2600$K. This makes the upper limit on $L_{EUV}/L_B$
independent of the effective temperature of the star.
 As $S_{eff}$ increases with the
increase in $T_{eff}$, the value of $S_{eff}$ corresponding
to the actual $T_{eff}$ of the star would only tighten the limit. It is
worth mentioning here that according to the relationships provided by
\cite{koppa13a}, $S_{eff}$ increases only by 2-3 times with the increase
in $T_{eff}$ from 2600 K to 7200 K. 

Taking $T_0=254$ K-  the equilibrium temperature of the Earth,
$R_P=6.378\times10^8$ cm - radius of the Earth,
 $M_P=5.9726\times10^{27}$ gm - mass of the Earth,
and $\epsilon=0.15$ \citep{watson81}, I derive, from equation~(\ref{result}),
 the upper limit on $L_{EUV}/L_B$ for planets within the outer boundary of
the habitable zone. The values are presented in Table~1.   
For the same set of parameters the numerical solution
of equation~(\ref{watson1}) and equation~(\ref{watson2}) for $q=0.7$ is also
presented in Table~1.

\begin{table}[h]
\begin{center}
\caption{Upper limit on $L_{EUV}/L_B$ for $S_{eff}$ corresponding to $T_{eff}=
2600$K under atmospheric conditions that determines the outer boundary of
the habitable zone. For $q=0.7$ the results are obtained numerically.}
\begin{tabular}{ccccc}
\hline \hline 
Conditions &  $S_{eff}$ & & $L_{EUB}/L_B$ & \\
&  & q=1.0 &  & q=0.7 \\
\hline 
Maximum Greenhouse & 0.215 & $1.33\times10^{-4}$ & & $1.56\times10^{-4}$ \\
Early Mars & 0.199 & $1.44\times10^{-4}$ & & $1.69\times10^{-4}$ \\
\hline
\end{tabular}
\end{center}
\end{table}

The critical EUV flux $S_c$ is less than 10 erg cm$^{-2}$ s$^{-1}$ which ensures
energy-limited escape. Also, equation (\ref{sol1}) gives 
the expanded radius $r_1=1.57R_P$ for $T_0=254$K. 

\section{DISCUSSIONS AND CONCLUSIONS}

   Estimating the stellar EUV luminosities is difficult not only
because the energy gets completely absorbed at the uppermost
layer of the Earth's atmosphere but also due to photoelectric
absorption by the interstellar medium  along the line of sight.
Using the ROSAT space telescope, \cite{hodgkin94} estimated
the EUV luminosities of a large number of nearby stars with 
spectral type ranging from F to M . Here, I use the bolometric
and EUV luminosities of planet-hosting stars of spectral type
ranging from A to M, given by \cite{sanz11} who
derived the EUV and bolometric luminosities 
by using the data from ROSAT,
XMM-Newton and Chandra space telescopes. I have considered
only those stars that are older than 1 Gyr.
It is found that out of all such stars, only two of
 the planet-hosting M stars satisfies the
criteria presented in this work. The value of
$L_{EUV}/L_{B}$ for 2MASS 1207 (spectral type M8) is
$1.23\times 10^{-3}$. For GJ 317 (M3.5), GJ 674
(M2.5) and for GJ 176 (M2.5V), the values of $L_{EUV}/L_B$ are 
$1.78\times10^{-3}$, $2.95\times10^{-4}$, $4.26\times10^{-4}$ respectively. 
Therefore, none
of these M stars can have a planet with habitable environment.
GJ 832 (M1.5V) has $L_{EUV}/L_B=10^{-4}$ and so it marginally satisfies
the habitability criteria. However, since the effective temperature of this
M1.5V spectral type star is higher than 2600K, this planet should also
be considered uninhabitable. 
On the other hand the values of $L_{EUV}/L_B$ for
GJ 436 (M2.5) and GJ 876 (M4V) are $1.58\times 10^{-5}$ and
$1.9\times 10^{-5}$ respectively. 
Therefore, these two M stars satisfies the EUV habitability 
criteria provided in Table~1.
This is expected because M stars are the faintest among
stars of all spectral types. As a consequence the HZ
of M stars is located very near to the stars and hence
a planet in the habitable zone is exposed to strong EUV radiation. 
It is worth mentioning here that a rocky planet in the habitable zone
of M stars may lose much of its volatiles during the formation 
because the pre-main sequence phase of such stars
are comparatively longer. However, if the planet accretes 
sufficiently large amount of water during formation or if it
were formed far away from the star and then migrated to the HZ,
it may remain habitable \citep{lissa07}. But strong EUV 
irradiation should make it uninhabitable.  

However, planet hosting stars of all other spectral types listed
by \cite{sanz11} have $L_{EUV}/L_B$ much lower than
the upper limits provided here.
The values of $L_{EUV}/L_B$ for GJ 86 (K1V) is $9.77\times 10^{-5}$. Therefore,
according to the present upper limit, this star marginally satisfies the
habitability criteria of having sufficient amount of water.
On the other hand, the values of $L_{EUV}/L_B$
 for HD 87883 (K0V), $\epsilon$
Eridani (K2V), HD 46375 (K1V), HD 93083 (K3V),
HD 130322 (K0V), HD 189733 (K1-K2) and HD 218566 (K3V) are
$2.75\times 10^{-5}$, $2.19\times 10^{-5}$, 
$1.02\times 10^{-5}$, $1.479\times10^{-5}$,
$1.95\times 10^{-5}$, $2.4\times 10^{-5}$ and 
$1.66\times 10 ^{-5}$ respectively. So, all  of them 
satisfies the upper limit for $L_{EUV}/L_B$.

  Hence, the upper limit
presented here is important in deciding the habitability of
planets around M  stars. Note that the numerical values of
the parameters involved in the present habitability condition do
not differ much from that of the Earth. The density of a rocky
planet should be about 4.0-5.5 gmcm$^{-3}$. Therefore, the
radius of a rocky planet does not exceed 1.6 times the Earth's
radius and the mass must be less than 5 times the Earth's mass \citep{rogers15}.
More massive planets are Neptune and Jupiter types of gaseous
planets \citep{marcy14}. Therefore, a large value of
$L_{EUV}/L_B$ for M stars may rule out the presence 
of habitable planets around them.

Owing to our poor knowledge on the actual environment of 
exoplanets, the habitability of rocky planets around the habitable zone
remains ambiguous and incomplete. However, we must impose as many
conditions as can  be determined directly and easily in order
to narrow down our search for planets that may possibly harbor
life. The present result provides such an important condition as it
depends on two observable quantities of the planet hosting star -
the bolometric and EUV luminosities. 
Spectra of habitable planets around M-type stars 
that do not satisfy the EUV criteria presented here
should show lack of hydrogen, nitrogen and oxygen in their
atmosphere and hence can confirm  the limit presented here.

I thank the reviewer for a critical reading of the manuscript and for
providing useful suggestions and comments.

\clearpage

\begin{figure*}
\begin{center}
\includegraphics[angle=0,width=14.0cm]{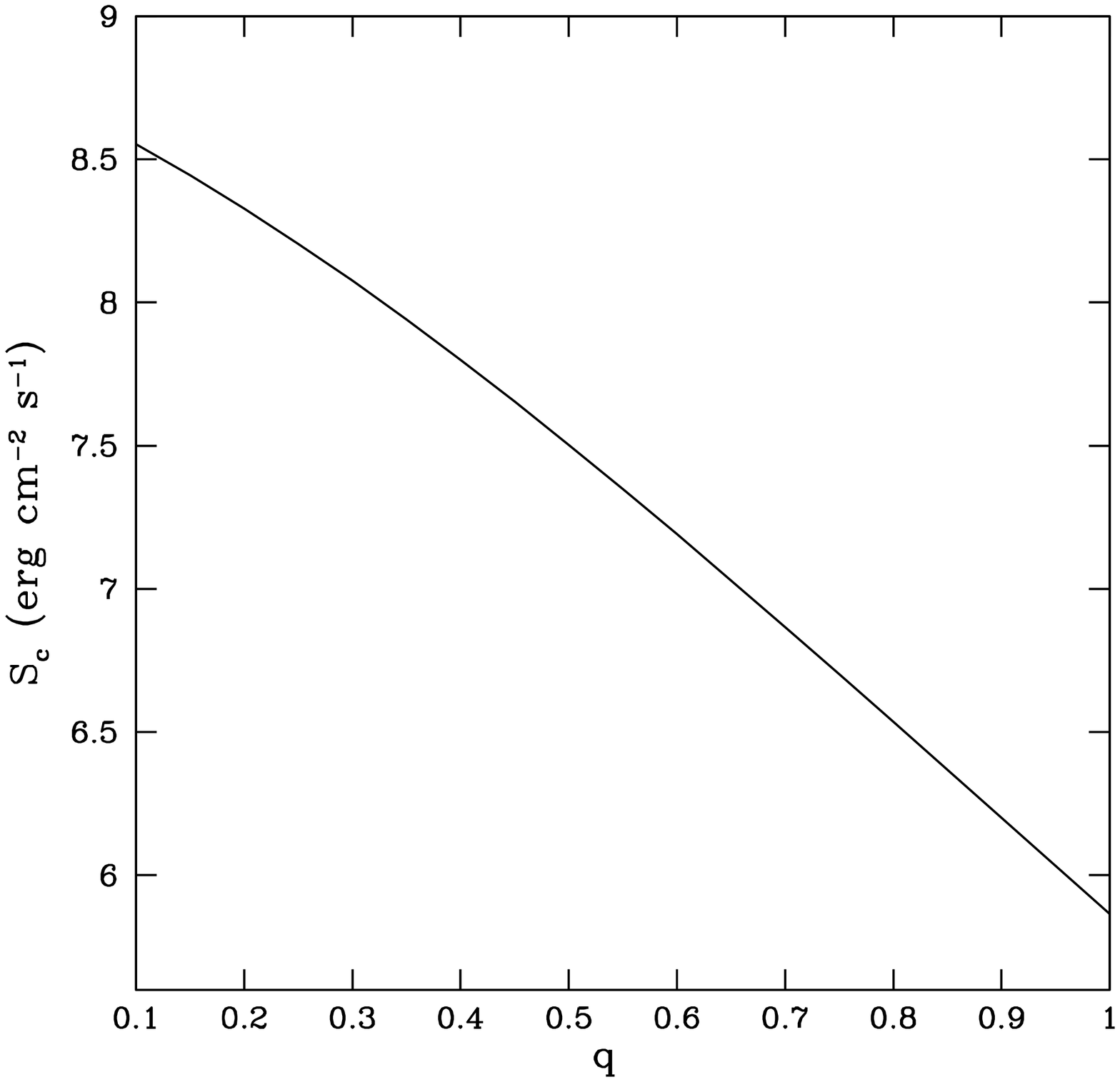}
\caption{The sphere-averaged,
efficiency-corrected and energy-limited critical EUV flux $S_c$
as a function of the conductivity index $q$.
Terrestrial parameters are used to calculate $S_c$.
For neutral gas, $q\simeq0.7$.}
\label{fig1}
\end{center}
\end{figure*}
\clearpage
\begin{figure*}
\begin{center}
\includegraphics[angle=0,width=14.0cm]{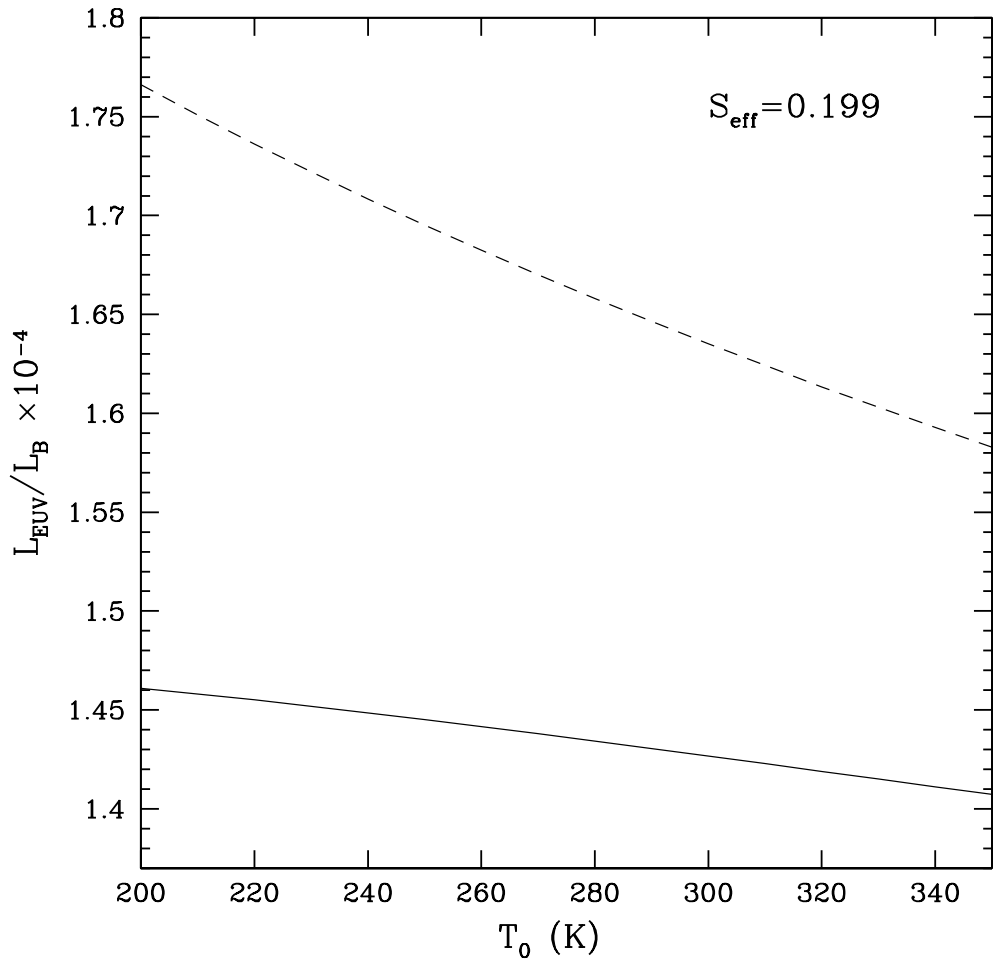}
\caption{$L_{EUV}/L_B$ for different values of $T_0$. Solid line represents
 $L_{EUV}/L_B$ for $q=1.0$ and dashed line represents that for $q=0.7$.} 
\label{fig2}
\end{center}
\end{figure*}

\end{document}